\documentclass[onecolumn, floatfix, preprintnumbers, eqsecnum, letterpaper, superscriptaddress, nofootinbib]{revtex4}
\usepackage{graphicx}
\usepackage{microtype}
\usepackage{amsmath}
\usepackage{amssymb}
\usepackage{subfigure}
\usepackage{hyperref}
\usepackage{url}
\usepackage{xcolor}
\usepackage{color}
\usepackage{mathrsfs}
\usepackage{calrsfs}
\usepackage{amsfonts}
\usepackage{eufrak}
\usepackage{tabularx}
\usepackage{eucal}
\usepackage{latexsym}
\usepackage{ragged2e}
\usepackage{epsfig}
\usepackage{textcomp}
\usepackage{inputenc}
\usepackage{caption}
\usepackage{subfigure}

\usepackage{marvosym}
\usepackage{ifsym}

\usepackage{lipsum}

\DeclareCaptionJustification{justified}{\leftskip=0pt \rightskip=0pt \parfillskip=0pt plus 1fil}
\definecolor{vividviolet}{rgb}{0.62, 0.0, 1.0}
\definecolor{amaranth}{rgb}{0.9, 0.17, 0.31}
\definecolor{palatinateblue}{rgb}{0.15, 0.23, 0.89}
\definecolor{brightpink}{rgb}{1.0, 0.0, 0.5}
\definecolor{cornflowerblue}{rgb}{0.39, 0.58, 0.93}
\definecolor{deepcarminepink}{rgb}{0.94, 0.19, 0.22}
\definecolor{radicalred}{rgb}{1.0, 0.21, 0.37}

\hypersetup{ linktoc=all,
	colorlinks, linkcolor={palatinateblue},
	citecolor={brightpink}, urlcolor={amaranth}
}

\graphicspath{{Images/}}

\graphicspath{{Images/}}

\def\@fnsymbol#1{\ensuremath{\ifcase#1\or \ddagger \or  $\textleaf$  \or \dagger
		\else\@ctrerr\fi}}%

\makeatother

\def\sideremark#1{\ifvmode\leavevmode\fi\vadjust{\vbox to0pt{\vss
			\hbox to 0pt{\hskip\hsize\hskip1em
				\vbox{\hsize1.3cm\tiny\raggedright\pretolerance10000
					\noindent #1\hfill}\hss}\vbox to8pt{\vfil}\vss}}}%
\def\beq{\begin{equation}}
	\def\eeq{\end{equation}}

\begin{document}
\title{Superradiance of rotating black holes surrounded by dark matter}

\author{Quan-Xu Liu}\email{quanxuliu@nuaa.edu.cn}
\address{College of Physics, Nanjing University of Aeronautics and Astronautics, Nanjing 211106, China}

\author{Ya-Peng Hu}\email{huyp@nuaa.edu.cn }
\address{College of Physics, Nanjing University of Aeronautics and Astronautics, Nanjing 211106, China}
\address{Key Laboratory of Aerospace Information Materials and Physics (NUAA), MIIT, Nanjing 211106, China}

\author{Tao-Tao Sui}\email{suitt14@lzu.edu.cn}
\address{College of Physics, Nanjing University of Aeronautics and Astronautics, Nanjing 211106, China}

\author{Yu-Sen An}\email{anyusen@nuaa.edu.cn \quad (Corresponding Author)}
\address{College of Physics, Nanjing University of Aeronautics and Astronautics, Nanjing 211106, China}

\begin{abstract}
In rotating black hole background surrounded by dark matter, we investigated the super-radiant phenomenon of massive scalar field and its associated instability.Using the method of asymptotic matching, we computed the amplification factor of scalar wave scattering to assess the strength of super-radiance. We discussed the influence of dark matter density on amplification factor in this black hole background. Our result indicates that the presence of dark matter has suppressive influence on black hole super-radiance.  We also computed the net extracted energy to further support this result. Finally, we analyzed the super-radiant instability caused by massive scalar field using the black hole bomb mechanism and found that the presence of dark matter has no influence on the super-radiant instability condition.


\end{abstract}

\maketitle



\section{Introduction}
With the discovery of gravitational waves and the release of the images of super-massive black holes in M87 and Sagittarius A* (Sgr A*)~\cite{LIGOScientific:2016aoc,LIGOScientific:2019fpa,EventHorizonTelescope:2019ths,EventHorizonTelescope:2019dse,EventHorizonTelescope:2022wkp,EventHorizonTelescope:2022wok}, the existence of black holes as real celestial objects in the universe has gradually been confirmed. The work by Penrose and Misner etal ~\cite{Penrose:1969pc,Penrose:1971uk,Misner:1972kx}suggested that black hole may be a huge source for energy supply in the future.Thus to truly utilize black holes, we need to continuously pay attention to its energy extraction problem. Various energy extraction methods from black hole have been proposed such as Penrose process \cite{Penrose:1971uk},magnetic reconnection \cite{Comisso:2020ykg,Khodadi:2022dff,Wei:2022jbi,Khodadi:2023juk} and superradiance~\cite{Misner:1972kx,Brito:2015oca}. Among these theories, the superradiance, as a general phenomenon in nature and an efficient method to extract energy from black hole, has been widely studied.  The reader can consult the comprehensive review \cite{Brito:2015oca} to gain a broad view of the study of superradiance.

For rotating black holes, super-radiance is actually the field theory version of the Penrose process. The Penrose process describes particles being captured by a Kerr black hole's ergo-sphere. When seen from distant observer, the particles in black hole ergo-sphere appear to gain negative energy, while other particles are ejected with higher energy than before because total energy is conserved. Thus effectively, we can extract energy from Kerr black hole~\cite{Penrose:1969pc,Penrose:1971uk}. Based on this work, instead of particles, Misner considered waves and transformed the Penrose process into a super-radiant scattering process of waves~\cite{Misner:1972kx}. He derived the condition of frequency for super-radiance to occur. After that, Teukolsky proposed that as long as some general conditions are satisfied, any bosonic fields (such as electromagnetic waves and gravitational waves) can always undergo this process~\cite{Teukolsky:1973ha}.\footnote{While interestingly, the superradiance is absent for fermionic field \cite{Brito:2015oca,Dai:2023zcj}}. 

It should be noted that superradiance is not only a feature of general relativity, it also appears in very broad gravitational theories. For discussions on superradiance phenomenon in various different gravitational theories, see for example Ref.\cite{Rosa:2020uoi,Chakravarti:2023wlc,Khodadi:2020cht,Franzin:2021kvj,Khodadi:2021owg,Alexander:2022avt,Jha:2022nzd,Khodadi:2022dyi,Jha:2022tdl,Yang:2022uze} where it was comprehensively shown that the super-radiance behavior is a useful tool to distinguish different gravitational theories and probe new physics beyond general relativity which may capture some quantum gravity signature. Furthermore,besides changing gravitational action, the superradiance phenomenon can also differ when consider adding some kinds of matter field to the asymptotic-flat vaccum solution as shown in Ref.\cite{Yang:2022uze,Chen:2021zqs,Cuadros-Melgar:2021sjy,Khodadi:2021mct,Richarte:2021fbi}. The discussions of superradiance are also not only limited to asymptotically flat spacetime, there are also generalizations of super-radiant phenomenon of black hole to de-Sitter and anti-de Sitter case \cite{Mascher:2022pku,Ishii:2022lwc}. 

Actually, in essence, super-radiance is the phenomenon where waves are scattered and amplified by a dissipative rotating object. Therefore, in any scattering scenario involving a dissipative rotating object, in-going waves have the possibility to undergo super-radiance, which is also not limited to black hole system.  In the case of horizon-less objects like stars, the presence of viscous material can also provide the necessary dissipation to trigger superradiance~\cite{Richartz:2013unq,Cardoso:2015zqa,Glampedakis:2013jya},which is very different from Hawking radiation \cite{Dai:2023ewf}. Thus it can been seen that superradiance is a very general phenomenon in nature which needs to be observed and utilized.

The existence of superradiance can lead to various interesting phenomenons, such as the instability of black holes~\cite{Witek:2012tr}. If the amplified waves from the super-radiant process encounter a “ mirror ” during the outgoing process\footnote{The mirror can be either natural (such as a massive scalar field or an AdS spacetime) or artificial~\cite{Furuhashi:2004jk,Cardoso:2006wa,Dolan:2007mj,Hod:2012zza,Dolan:2012yt,Zhu:2014sya,Green:2015kur,Huang:2018qdl,Destounis:2019hca,Huang:2019xbu,Li:2019tns,Xu:2020fgq,Vieira:2021nha}}, the waves are reflected again, resulting in secondary superradiant amplification. This process continues iteratively, and if the superradiant mode is confined near the black hole, it undergoes exponential growth. This growth will disrupt the equilibrium, lead to instability and cause the radiation to jet outward like a bomb. This mechanism is well known as the “ black hole bomb ”~\cite{Berti:2019wnn,Press:1972zz,Cardoso:2004nk,Cardoso:2013krh,Herdeiro:2013pia,Dolan:2015dha,Dias:2018zjg}. Certainly, the instability arising from superradiance presents interesting theoretical implications. For example, it can give rise to new black hole solutions that violate the no-hair theorem~\cite{Herdeiro:2016tmi,Herdeiro:2017phl,Degollado:2018ypf,Herdeiro:2020xmb,Rahmani:2020vvv}. Moreover, superradiance can also impose constraints on ultra-light bosonic particles beyond the standard model~\cite{Brito:2014wla,Correa:2024xki} which makes it a promising natural laboratory for particle detection in high-energy physics.

Among these interesting directions related to superradiance, in this work, we are interested in one aspect, which is combining the black hole superradiance with the dark matter.

The motivations come from two sides. On the one hand, with the development of astronomy, an increasing number of theoretical predictions related to black holes are going to be confirmed in astronomical observations in the future. Thus black hole may serve as a new ground for detecting new physics. Based on this point,one interesting direction will be the detection of dark matter and dark energy, which is confirmed to be ubiquitous in our universe by many indirect evidence in cosmology~\cite{Rubin:1980zd,Persic:1995ru,Bertone:2016nfn}. Although dark matter is challenging to be directly detected due to its lack of electromagnetic interactions, a wealth of observational data indirectly suggests that typical galaxies are filled with abundant dark matter~\cite{Rubin:1980zd,Persic:1995ru,Bertone:2016nfn}. These observations rely on the fact that dark matter participate in gravitational interactions, therefore, as an gravitational objects, black hole may be the next avenue for us to detect the evidence of dark matter.

On the other hand, the establishment of black hole no-hair theorem in asymptotic flat spacetime only considers electro-magnetic field, which results in limited black hole solution such as Reissner-Nordstrom black hole and Kerr-Newman black hole. When taking into account the presence of dark matter, certainly we can get more interesting black hole solutions which will bring new physics of black holes in asymptotically flat spacetime. 

To consider the backreaction of dark matter on the spacetime and get new black hole solution, it is necessary to know the dark matter energy profile. There are no consensus on the dark matter energy profile while several forms have been proposed using numerical method, such as Hernquist type \cite{Hernquist:1990be} and Navarro-Frenk-White type \cite{Navarro:1996gj}.It is important to know the back-reaction caused by these dark matter profiles, but it is very hard to get analytical black hole solutions without other assumptions. The lack of analytic solution will hinder further studies on the dark matter black hole. For current progress on this direction, the reader can consult Ref.\cite{Shen:2023erj}, while this paper is still based on dark matter profiles different from Hernquist and NFW type. Based on this point, one way to go is to loose the constraint about the dark matter energy profile to get analytical black hole solutions and then analytically investigate the physical phenomenon therein to compare with observations. 

In such way, researchers have discovered various spherically symmetric solutions in the presence of dark matter. The first such solution was derived by Kiselev for quintessential dark matter \cite{Kiselev:2002dx,Kiselev:2003ah}. After that, by combining dark matter with phantom dark energy, Li etal found another black hole solution in the presence of dark matter which is so called "perfect fluid dark matter (PFDM)" solution~\cite{Li:2012zx}. After that it has also been further extended to the rotating~\cite{Toshmatov:2015npp,Xu:2017bpz} and charged rotating cases~\cite{Das:2020yxw} by using Newman-Janis algorithm. These solutions are useful to understand galactic rotation curve as discussed in \cite{Kiselev:2002dx,Kiselev:2003ah,Li:2012zx},thus can be seen as black hole with dark matter. These solutions provide a good background to investigate the effect of dark matter on black hole shadow \cite{Haroon:2018ryd},quasi-normal mode \cite{Jusufi:2019ltj} and black hole thermodynamics \cite{Liang:2023jrj}. But many aspects are still poorly explored. As black hole superradiance has significant observational meaning and close relation to other observations such as black hole shadow and photon ring \cite{Roy:2021uye,Chen:2022nbb}.Considering super-radiance phenomenon in this black hole background which is surrounded by dark matter would be an intriguing research direction both in theoretical and observational sense. So in this work we will focus on exploring the superradiant amplification effects of rotating black holes immersed in perfect fluid dark matter. 

The structure of this paper is as follows:
In the Sec~\ref{section2}, we provide a brief introduction to the solutions of rotating black holes in the presence of PFDM.
In the Sec~\ref{section3}, we first discuss the conditions for the occurrence of superradiance with a massive scalar field. Then, we use a semi-analytical method to calculate the superradiant amplification factor. In the Sec~\ref{energy}, we explore the extraction of energy from the PFDM rotating black hole using massless scalar field.
In the Sec~\ref{section4}, we analyze the instability of black hole superradiance in the presence of a massive scalar field by examining the effective potential.
The Sec~\ref{section5}, we conclude the paper and give an explanation of our mean results from black hole spacetime structure perspective. This research will help to understand the effect of dark matter on black hole physics. During this work, we use geometric units where $G=c=1$.
\section{Rotating black hole surrounded by perfect fluid dark matter}
\label{section2}
In this section, we will briefly introduce the rotating black hole solution immersed in dark matter, which is the background throughout our work. The action of Einstein gravity in the presence of dark matter is written as~\cite{Das:2020yxw}:
\begin{align}
S=\int d^4x\sqrt{-g}\left[\dfrac{\bar{R}}{16 \pi}+ \mathcal{L}_{\text{DM}}+\mathcal{L}_{\text{DE}}\right],\label{darkmatteraction}
\end{align}
where $g=det(g_{\mu\nu})$ is the determinant of the metric tensor, $\bar{R}$ is the Ricci scalar, 
and $\mathcal{L}_{\text{DM}}$ denotes the dark matter Lagrangian,$\mathcal{L}_{\text{DE}}$ denotes the contribution of dark energy, which is taken to be phantom dark energy in \cite{Li:2012zx}.We do not write explicit form of dark energy Lagrangian as we will not use it throughout the paper. We also neglected the contribution of electric charge as it is not supported astrophysically.

Varying this action with respect to metric, the Einstein equation is obtained as follows
\begin{align}
\bar{R}_{\mu\nu}-\dfrac{1}{2}g_{\mu\nu}\bar{R}&=8\pi\ T_{\mu\nu}^{DS},\label{DMEinsteinequation}
\end{align}
where $T^{DS}_{\mu\nu}$ is the energy-momentum tensor corresponding to the dark sector(dark matter plus dark energy). We choose the perfect fluid stress tensor for the dark matter
\begin{align}
(T^{\mu}~_{\nu})^{DM}=\ {\rm diag}\left(-\rho_{DM},0,0,0\right),\label{DMtensor}
\end{align}
considering the involvement of dark energy (phantom dark energy in \cite{Li:2012zx}), the final energy-momentum tensor for dark sector takes the following form
\begin{align}
(T^{\mu}~_{\nu})^{DS}=\ {\rm diag}\left(-\rho,P_r,P_\theta,P_\phi\right),\label{PFDMtensor}
\end{align}
where $\rho=\rho_{DM}+\rho_{DE}$ is the total energy density and $P_i$ is the pressure of dark energy. Strictly speaking,stress tensor $(T^{\mu}~_{\nu})^{DS}$ in Eq.(\ref{PFDMtensor}) actually contains dark matter plus phantom dark energy which is not a perfect fluid nor a matter \footnote{The perfect fluid only refers to the stress energy tensor of dark matter(\ref{DMtensor}), when taking into account the dark energy, the total stress energy tensor (\ref{PFDMtensor}) no longer has the feature of perfect fluid,while many works discussing "PFDM" black hole describe little about the associated dark energy , which causes some confusions.}, while we still abbreviate this as dark matter throughout this paper to match the notation of previous work. Moreover, for further details about the properties of dark energy therein , the readers can consult Ref.~\cite{Das:2020yxw}. 

By further requiring the stress tensor $(T^{\mu}~_{\nu})^{DS}$  having following condition: 
\begin{equation}
    \rho=-P_r, \quad P_\theta=P_\phi=\frac{\rho}{2}
\end{equation}
A static spherically symmetric solutions can be obtained  (\ref{DMEinsteinequation}) as
\begin{align}
ds^2=-f(r)dt^2+\dfrac{1}{f(r)}dr^2+r^2(d\theta^2+\sin^2\theta d\phi^2)\label{ansatzmetric},
\end{align}
with
\begin{align}
f(r)=1-\dfrac{2M}{r}+\dfrac{\lambda'}{r}\ln\left(\dfrac{r}{|\lambda'|}\right).\label{RNdSPFDMmetric}
\end{align}
Here, $M$ is the mass parameter of the black hole, $Q$ is the charge of black hole and $\lambda'$ is the PFDM parameter. From the Einstein equation, we can get 
\begin{equation}\label{rholabel}
    \rho=-\frac{\lambda'}{8\pi r^3},
\end{equation}
thus parameter $\lambda'$ denotes the contribution of dark sector to the total energy density which is why we call it PFDM parameter. Invoking the weak energy condition $\rho \geqslant 0$, we can find that $\lambda'\leqslant 0$ and in the following, we will remove the absolute value symbol of $\lambda'$ from the equation (\ref{RNdSPFDMmetric}) and redefine $-\lambda'$ as $\lambda$ ~\cite{Liang:2023jrj},
\begin{align}
f(r)=1-\dfrac{2M}{r}-\dfrac{\lambda}{r}\ln\left(\dfrac{r}{\lambda}\right),\label{RNdSPFDMmetric}
\end{align}
where $\lambda$ satisfies $\lambda>0$. It can be easily seen that at asymptotic infinity $r\to \infty$, for the last term, the limit $\lim_{r\to \infty} \frac{\lambda}{r}\ln(\frac{r}{\lambda})$ vanishes, thus this solution is an asymptotic flat solution.\footnote{Note that this metric is different from Kiselev black hole solution \cite{Kiselev:2002dx,Kiselev:2003ah} where quintessential dark matter is taken into account. For Kiselev black hole $f(r)=1-\frac{2M}{r}-(\frac{r_{n}}{r})^{3w_{n}+1}$ which is different from Eq.(\ref{RNdSPFDMmetric}), $w_{n}$ is the state parameter satisfying $-1<w_{n}<0$ which reflects the quintessential property. For $w_{n}<-1/3$, Kiselev black hole is not asymptotically flat.} We need to stress here that as was shown in Ref.\cite{Liang:2023jrj} from black hole thermodynamics, actually the PFDM surrounded black hole metric can not be valid all across the spacetime, it is only valid up to a halo region $r<r_{halo}$. But unlike thermodynamics \cite{Liang:2023jrj},for the computation of superradiance, we do not need to consider the asymptotic infinity. We also do not need to know the detailed properties of $r_{halo}$ but only need to assume $r_{halo}-r_{h}\gg M$ in order to perform the "analytic asymptotic matching" algorithm when computing superradiance. This assumption is supported by the observation of dark matter halo.

Realistic black holes in our universe always rotate, so the black hole with spin should be taken into account.Newman-Janis algorithm is the universal method to obtain the rotating black hole metric from the static spherical symmetric solution \cite{Toshmatov:2015npp,Xu:2017bpz,Azreg-Ainou:2014aqa,Azreg-Ainou:2014nra,Azreg-Ainou:2014pra}, the resulting rotating PFDM black hole solution reads




\begin{align}
     ds^2=&-\dfrac{1}{\rho^2}(\Delta-a^2\operatorname{sin}^2\theta)dt^2+\dfrac{\rho^2}{\Delta}dr^2+\rho^2d\theta^2-\dfrac{2a\operatorname{sin}^2\theta}{\rho^2}(2Mr+\lambda r\ln(\dfrac{r}{\lambda}))dtd\phi \notag\\&+\operatorname{sin}^2\theta(r^2+a^2+\dfrac{a^2\operatorname{sin}^2\theta}{\rho^2}(2Mr+\lambda r\ln(\dfrac{r}{\lambda})))d\phi^2,\label{newmetric}
\end{align}
with 
\begin{align}
  & \Delta=r^2+a^2-2Mr-\lambda r\ln({\dfrac{r}{\lambda}}), \notag\\ & \rho^2=r^2+a^2\operatorname{cos}^2\theta.\label{newdelta}
\end{align}
The positions of the event horizon($r_h$) and the Cauchy horizon($r_c$) are obtained from the solution of $\Delta=0$ , which can only be obtained numerically due to the presence of the logarithmic term.

Note that in this case, the stress energy distribution will change to
\begin{equation}
    \rho=-p_{r}=\frac{\lambda r}{8\pi(r^{2}+a^{2}\cos^{2}\theta)^{2}}
\end{equation}
\begin{equation}
    p_{\theta}=p_{\phi}=\frac{\lambda}{8\pi(r^{2}+a^{2}\cos^{2}\theta)^{2}}(r-\frac{r^{2}+a^{2}\cos^{2}\theta}{2r})
\end{equation}
to maintain this rotating solution. It can be easily seen that the expression of energy and pressure for spherical symmetric case (\ref{rholabel}) is recovered when $a=0$.

\begin{figure}[h]

\includegraphics[width=3.3in]{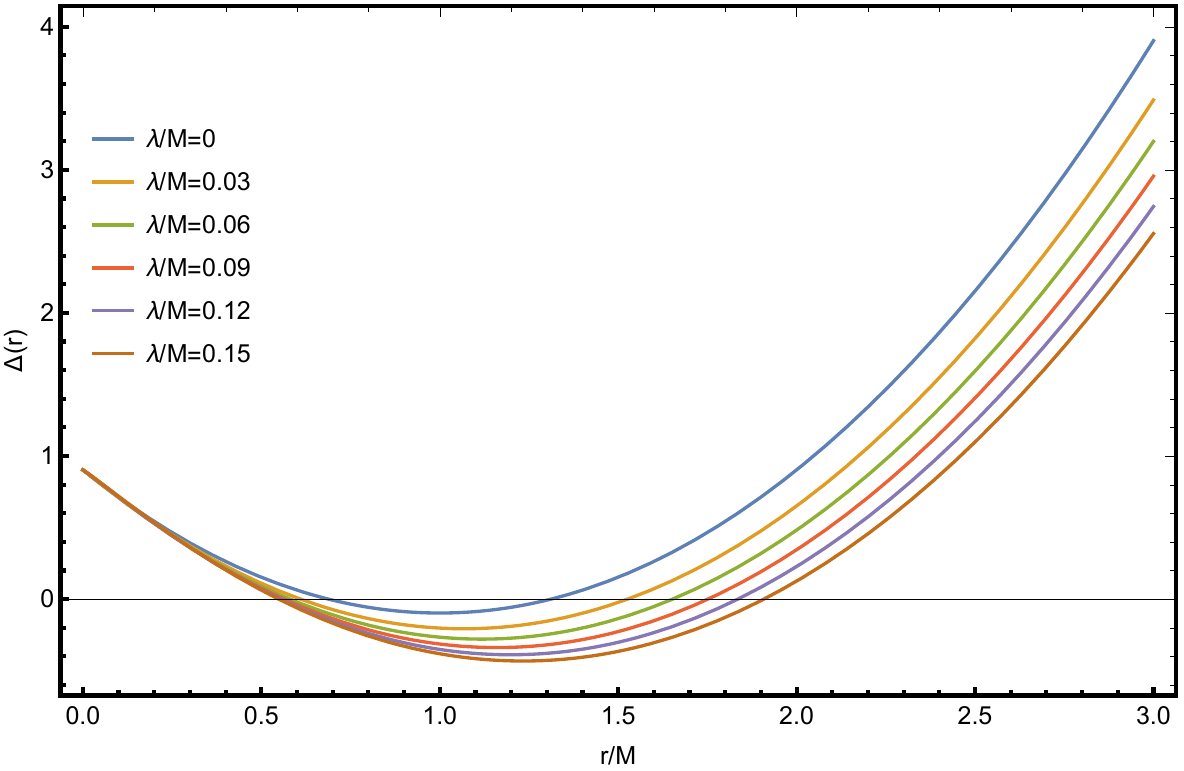}\label{delta1}
\hspace{2mm}
\includegraphics[width=3.3in]{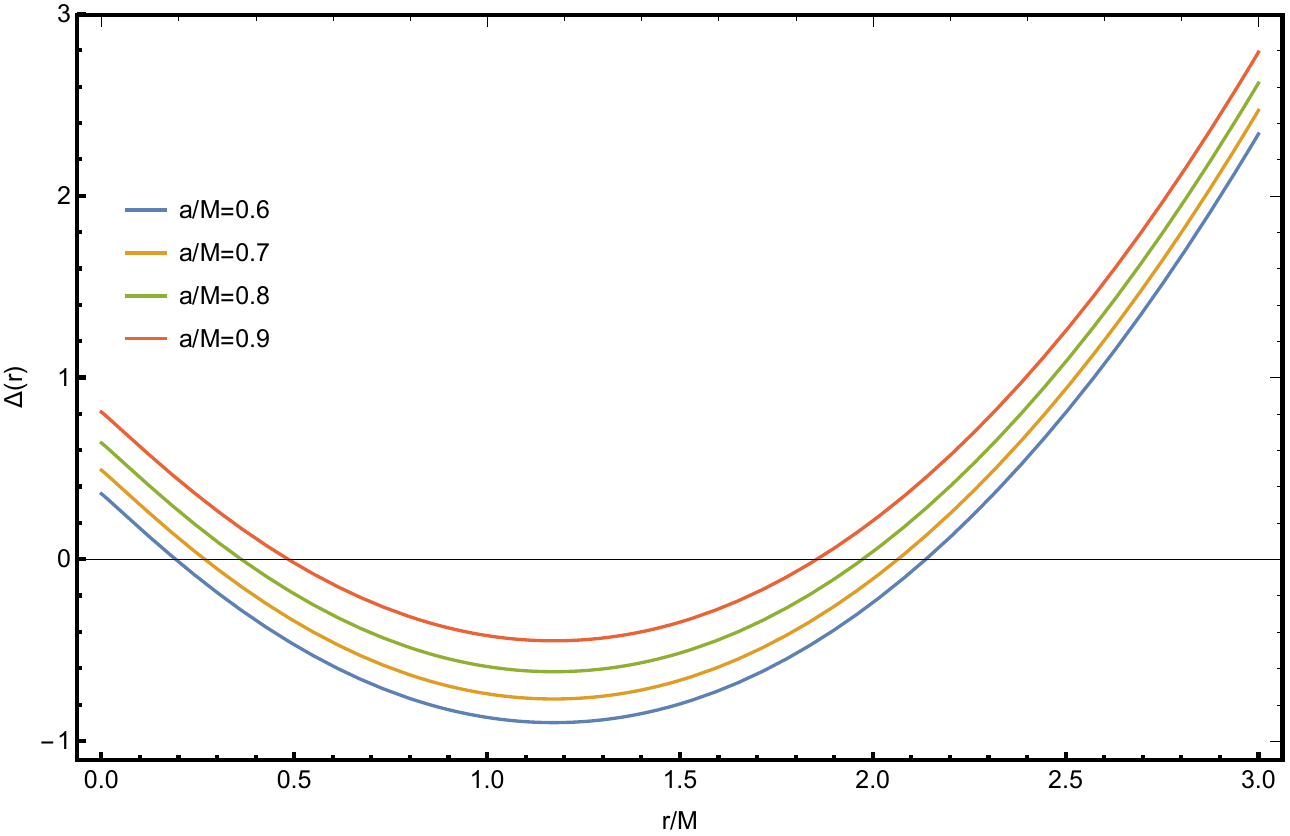}\label{delta2}

\caption{The horizon positions of rotating black holes with PFDM for different parameters, left panel: $a/M=0.95$, right panel: $\lambda/M=0.1$. The variations in the parameter $\lambda$ mainly affect the position of the outer horizon, while the rotation $a$ influences the position of both inner and outer horizons significantly. }
\label{delta}
\end{figure}

There are strong gravity constraints on the ratio between dark matter density and black hole mass parameter $\lambda/M$. It has been reported in Ref.\cite{Vagnozzi:2022moj} that the ratio has an upper bound $\lambda/M<0.15$ to $2\sigma$ confidence level by using the Event Horizon Telescope image of Sagittarius A. Thus for observational meaning, we also choose our parameter falls within this region. We plot the metric function for various parameters in Fig~\ref{delta}.From Fig ~\ref{delta},the black hole has two horizons and we can observe that different parameters $\lambda$ and $a$ lead to variations in the position of these two horizons. Compared to the standard Kerr black hole($\lambda/M=0$), the presence of dark matter causes the outer horizon of the black hole to be larger and the inner horizon to be smaller. The variations in the parameter $\lambda$ mainly affect the position of the outer horizon, while having smaller effect on the inner horizon. This is different from the effect caused by the black hole spin. As shown in right panel of Fig ~\ref{delta}, the spin $a$ influences the position of both inner and outer horizons significantly. 

\section{Dark matter influence on the superradiant amplification factor of massive scalar field }
\label{section3}
In this section, we consider the superradiance phenomenon of massive scalar field in a rotating black hole in the presence of PFDM. The equation of motion for a massive scalar field $\Phi$ is governed by the Klein-Gordon equation~\cite{Bezerra:2013iha}
\begin{align}
    \left(\nabla^a \nabla_a-\mu^2\right) \Phi=(\sqrt{-g})^{-1} \partial_\mu\left(\sqrt{-g} g^{\mu \nu} \partial_\nu \Phi\right)-\mu^2 \Phi=0,\label{kg}
\end{align}
where $\mu$ is the mass of the scalar field. 
We can rewrite (\ref{kg}) more explicitly in the Boyer-Lindquist coordinates as 
\begin{align}
    &\left(-a^2 \operatorname{sin}^2\theta+\frac{\left(a^2+r^2\right)^2}{\Delta}\right) \partial_{t}^{2} \Phi+\frac{2a\left(a^2+r^2-\Delta\right)}{\Delta} \partial_{t}\partial_{\phi} \Phi +\left(-\operatorname{csc}^2\theta+\frac{a^2}{\Delta}\right) \partial_{\phi}^{2} \Phi
    -\partial_r\left(\Delta \partial_r \Phi\right)-\frac{1}{\operatorname{sin}\theta} \partial_\theta\left(\operatorname{sin}\theta \partial_\theta \Phi\right) \notag\\&+\mu^2\left(r^2+a^2 \operatorname{cos}^2\theta\right) \Phi=0,\label{kgequation}
\end{align}
Since the background spacetime is stationary and axial-symmetric, we use following ansatz to separate the scalar perturbation~\cite{Konoplya:2018arm},
\begin{align}
\Phi\left(x^\mu\right)=e^{-i \omega t} e^{i m \phi} S_{l m}(\theta) R_{l m}(r),\label{ansatz}
\end{align}
Here, $S_{l m}(\theta)$ is spheroidal harmonics with $-m\leq l \leq m$ and $R_{l m}(r)$ is the radial part of the wave function. Substituting (\ref{ansatz}) to (\ref{kgequation}),Eq. (\ref{kgequation}) can be separated into
\begin{align}
    &\frac{d}{dr}\left(\Delta\frac{d}{dr}R_{lm}(r)\right)+\left(\frac{\left((a^2+r^2)\omega -a m\right)^2}{\Delta}-(r^2 \mu^2-2 a m \omega+a^2 \omega^2+\Lambda_{lm})\right ) R_{lm}(r)=0,\label{radialeq}
\end{align}
\begin{align}
&\frac{1}{\sin \theta} \frac{d}{d \theta}\left(\sin \theta \frac{d S_{l m}}{d \theta}\right) +\left(a^2\left(\omega^2-\mu^2\right) \cos ^2 \theta-\frac{m^2}{\sin ^2 \theta}+\Lambda_{l m}\right) S_{l m}(\theta)=0
,\label{angulareq}
\end{align}
where $\Lambda_{lm}$ is the eigenvalues of the angular equation. In the non-rotating limit $a^{2}\omega^{2}\ll 1$, the eigenvalue $\Lambda_{lm}$ approaches $l(l+1)$ .\footnote{As we consider the scalar particle, so $s=0$ throughout the paper} The only interesting part is the Eq.(\ref{radialeq}). By changing variable $U(r)=\sqrt{a^2+r^2} R_{lm}$, and introduce the tortoise coordinate $dr_*=\frac{r^2+a^2}{\Delta} dr$ . Then, the radial function (\ref{radialeq}) takes the following Schrodinger-type equation
\begin{align}
    \frac{d^2U(r_*)}{dr_*^2}+V(r) U(r_*)=0,\label{sde}
\end{align}
with the effective potential $V(r)$ given by
\begin{align}
 V(r)=&\left(\omega-\frac{a m}{a^2+r^2}\right)^2-\frac{\Delta\left(l(l+1)+r^2 \mu^2-2 m a \omega+a^2 \omega^2\right)}{\left(a^2+r^2\right)^2}\notag\\&
 -\frac{\Delta^2}{\left(a^2+r^2\right)^3}+\frac{3 r^2 \Delta^2}{\left(a^2+r^2\right)^4}-\frac{\Delta\left(-2 M r+2 r^2-\lambda r-\lambda r \ln \left(\frac{r}{\lambda}\right)\right)}{\left(a^2+r^2\right)^4}.\label{veff}  
\end{align}
Before we proceed, we would like to mention the scattering of the scalar field under the effective potential. The asymptotic behaviors of the effective potential at the event horizon and spatial infinity are as follows
\begin{align}
    \lim _{r \rightarrow r_{h}} V(r)=\left(\omega-m \Omega_{h}\right)^2 \equiv k_{h}^2, \quad \Omega_{h} \equiv \frac{a}{r_{h}^2+a^2},\label{kh}
\end{align}
\begin{align}
    \lim_{r \rightarrow \infty} V(r)=\omega^2-\lim_{r \rightarrow \infty} \frac{r^2 \mu^2 \Delta}{(r^2+a^2)^2}=\omega^2-\mu^2\equiv k_{\infty}^2,\label{kinfty}
\end{align}
where $\Omega_{h}$ is the angular velocity of the event horizon.

Then, we have the following asymptotic behaviour of the solutions (boundary conditions) for Eq.(\ref{sde})
\begin{align}
     & U(r_*)=T e^{-i k_h r_*},\quad r_*\rightarrow -\infty (r\rightarrow r_h),\notag\\ & 
    U(r_*)=I e^{-i k_\infty r_*}+Re^{i k_\infty r_*},\quad r_*\rightarrow \infty (r\rightarrow \infty).\label{boundary}
\end{align}
Thus the wave at horizon and asymptotic infinity becomes simple plane wave with different frequency. There is in-going wave at horizon with amplitude ${T}$, and at infinity, there are both in-going wave with an amplitude ${I}$ and a reflected outgoing wave with amplitude ${R}$. Now, by computing and equating the Wronskian quantity
 \begin{align}
     W=(U \frac{dU^*}{dr_*}-U^*\frac{dU}{dr_*})\label{wronskian}
 \end{align}
at the event horizon and spatial infinity we can get
\begin{align}
    \left|{R}\right|^2=\left|I\right|^2-\frac{\omega-m \Omega_h}{\sqrt{\omega^2-\mu^2}}\left|T\right|^2.\label{relation}
\end{align}
which implies that the amplitude of the reflected waves must be larger than the amplitude of the incident wave if the condition
\begin{align}
  \mu < \omega < m \Omega_h,\label{srcondition}
\end{align}
is obeyed and therefore, the super-radiant scattering occurs ($|R|^2 > |I|^2$) .
From Fig.\ref{delta} and expression of horizon angular velocity $\Omega_{h}=\frac{a}{r_{h}^{2}+a^{2}}$,as the event horizon radius $r_{h}$ increase as we increase dark matter density, the horizon angular velocity $\Omega_{h}$ decreases which results in the shrink of frequency domain of super-radiance.

Besides superradiance frequency, another important physical quantity is the magnitude of superradiance. To access it,  the amplification factor can be defined which reads
\begin{align}
    Z_{lm}=\frac{|R|^2}{|I|^2}-1.\label{Z}
\end{align}
To concretely compute $Z_{lm}$ , we must solve radial equation (\ref{radialeq}). However, the radial equation (\ref{radialeq}) is not exactly solvable analytically, thus we will take advantage of a semi-analytical method which is called “analytical asymptotic matching”~\cite{Starobinsky:1973aij}. The procedure of this approach is that, we first find two approximate solutions each valid over a partial range and then match the two solutions in the overlapping region. After matching,a single approximate solution can be obtained. In this paper, we call these two regions “near-region” and “far-region” which respectively, denote the region around event horizon ($r-r_h\ll \omega^{-1}$) and far away from the event horizon ($r-r_h\gg M$). And the two solutions have to be in an “overlapping region” to match where $M\ll r-r_h\ll \omega^{-1}$. 

By applying this method, we should introduce two assumptions: we should first take the low-frequency regime of the perturbation $a \omega \ll 1$ and we also assume BH's size is smaller than the Compton wavelength of the scalar field $\Phi$ i.e., $M \omega \ll 1$ (or $\mu M \ll 1$)~\cite{Detweiler:1980uk}. 

The radial equation (\ref{radialeq}) can be rewritten in the form 
\begin{align}
    &\Delta^2 \frac{d^2 R_{lm}(r)}{dr^2}+\Delta \frac{d\Delta}{dr} \frac{dR_{lm}(r)}{dr}+((\omega (a^2+r^2)-am)^2-\Delta(r^2\mu^2-2am\omega+a^2\omega^2+l(l+1)))R_{lm}(r)=0,\label{reradialeq}
\end{align}

\begin{figure}[h]
    \centering
    \includegraphics[width=0.8\linewidth]{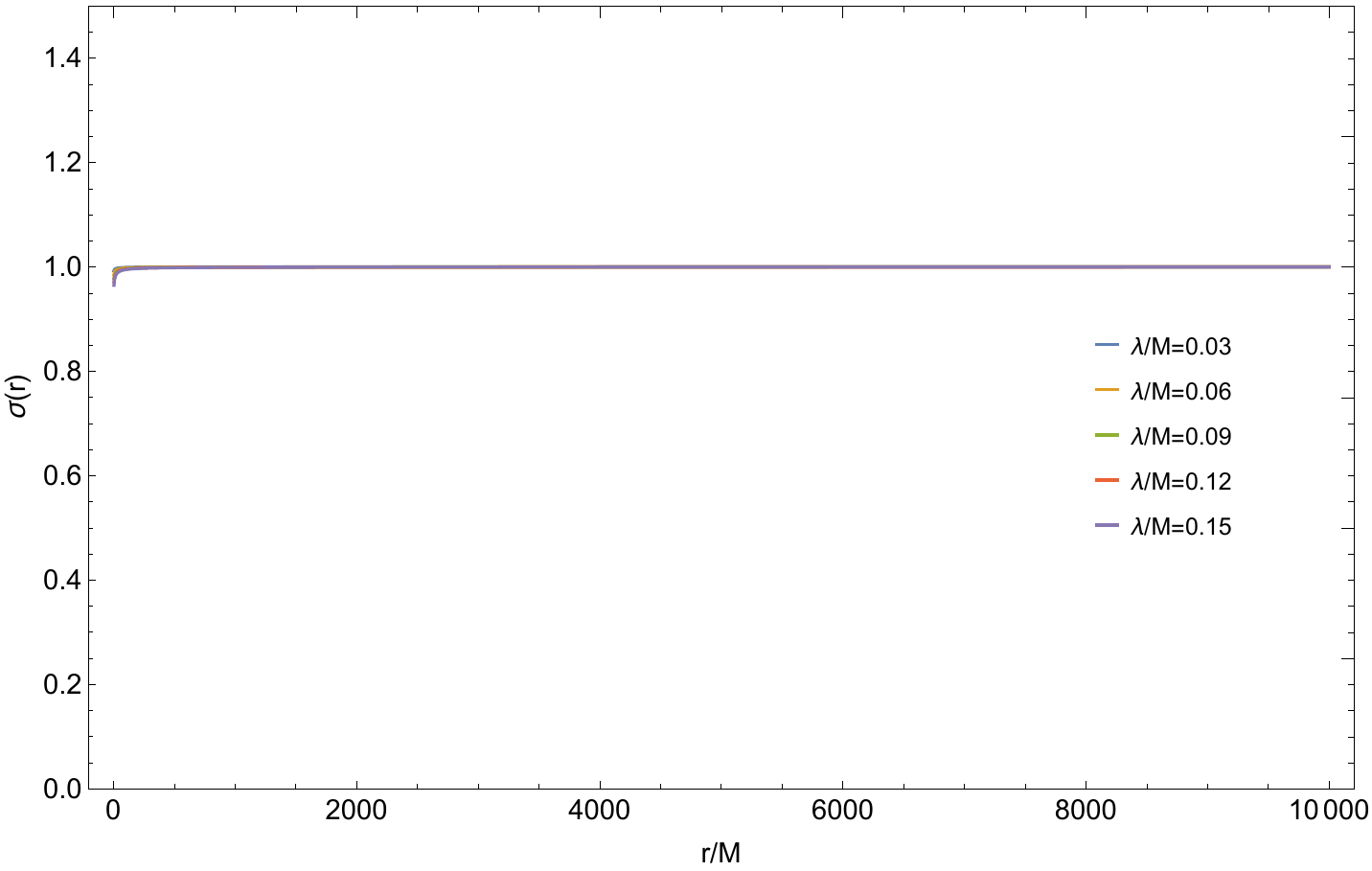}
    \caption{The numerical results of the proportional function $\sigma(r)$ with  $a/M=0.95$ and $\lambda/M=0.03$, $\lambda/M=0.06$, $\lambda/M=0.09$,$\lambda/M=0.12$ and $\lambda/M=0.15$.It can be seen that for all the range of $r$, $\sigma(r)$ approaches $1$ so we can approximate $\sigma$ to $1$ in the discussion. }
    \label{ff}
\end{figure}

In Kerr black hole, we know that $\Delta_{K}=r^2-2Mr+a^2=(r-r_h)(r-r_c)$.While in our PFDM case with two horizons, because of the log term, $\Delta$ in (\ref{reradialeq}) can only be expressed as $\Delta=r^2-2Mr+a^2-\lambda r \ln({\frac{r}{\lambda}})=(r-r_h)(r-r_c)\sigma(r)$ where $\sigma(r)$ should be a monotonic function with no zeros.
However, as plotted in Fig. \ref{ff}, at least in our interested parameter region($\lambda$ region), we can observe that $\sigma(r)\approx 1$ from horizon to infinity \footnote{There is small corrections for the region very close to the horizon ,so the approximation made in the near horizon region may get a little modification.}, so in order to solve the Eq.(\ref{reradialeq}) semi-analytically, we still approximate $\Delta$ as $(r-r_h)(r-r_c)$ in PFDM case without loss of much accuracy.Taking into account the correction caused by $\sigma(r)$ is interesting but difficult.It may need significant modification to the semi-analytical treatment of Teukolsky equation. But this correction is very small compared to the shift of $r_{h}$ and $r_{c}$ caused by dark matter profile, so we only focus on the leading order correction caused by dark matter in this work. 

By applying the change of variable $x=\frac{r-r_h}{r_h-r_c}$ and also using approximation $a\omega \ll 1$, the equation (\ref{reradialeq}) can be rewritten as

\begin{align}
 & x^{2}(1+x)^2 \frac{d^2 R_{l m}(x)}{d x^2}+x(x+1)(2 x+1) \frac{d R_{l m}(x)}{d x}\notag\\ & +\left(k^2 x^4-\left(\omega^2 a^2+l(l+1)\right) x(x+1)-\mu^2\left(\left(r_{h}-r_{c}\right) x+{r}_{h}\right)^2 x(x+1)+B^2\right) R_{l m}(x)=0,\label{changeode}
\end{align}
where
\begin{align}
x & =\frac{r-{r}_{h}}{{r}_{h}-{r}_{c}}, \\
k & =\omega\left({r}_{h}-{r}_{c}\right), \\
B & =\frac{{r}_{h}^2}{{r}_{h}-{r}_{c}}\left(\omega-m \Omega_h\right).
\end{align}
In the following, we will solve Eq.(\ref{changeode}) in the near horizon limit and far-region limit and use asymptotic matching method to compute the reflection coefficient.

\textbf{(i) Near-(horizon) region solution}\quad When approaching the event horizon, we have $kx\ll 1$ and $\mu^2((r_h-r_c)x +r_h)^2 \approx \mu^2 r_{h}^2 \ll 1$, so equation (\ref{changeode}) can be reduced to
\begin{align}
&x^2(x+1)^2 \frac{d^2 R_{l m}(x)}{d x^2}  +x(x+1)(2 x+1) \frac{d R_{l m}(x)}{d x}+\left(B^2-l(l+1) x(x+1)\right) R_{l m}(x)=0.\label{nearode}
\end{align}
The general solution of equation (\ref{nearode}) satisfying the boundary condition (\ref{boundary}) is given by the hypergeometric functions
\begin{align}
    R_{l m}(x)=C\left(\frac{x+1}{x}\right)^{i B} \times \quad _2 F_1(-l, l+1,1-2 i B,-x),\label{solnear}
\end{align}
where $_{2}F_{1}(-l, l+1,1-2 i B,-x)$ is hypergeometric function. In order to match the solutions, it is essential to know that the large $x$ behavior of above solution is 
\begin{align}
R_{l m}(x) & \sim  C x^l \frac{\Gamma(1-2 i B) \Gamma(2 l+1)}{\Gamma(1+l-2 i B) \Gamma(l+1)} +C x^{-l-1} \frac{\Gamma(1-2 i B) \Gamma(-2 l-1)}{\Gamma(-l) \Gamma(-l-2 i B)}.\label{nearfarsol}
\end{align}

\textbf{(ii) Far-region solution} \quad In this region we consider $x\rightarrow \infty$, the equation (\ref{changeode}) is approximately given by
\begin{align}
    \frac{d^2 R_{l m}(x)}{d x^2}+\frac{2}{x} \frac{d R_{l m}(x)}{d x}+\left(D^2-\frac{l(l+1)}{x^2}\right) R_{l m}(x)=0,\label{farode}
\end{align}
where $D=(r_h-r_c)\sqrt{\omega^2-\mu^2}$ . The general solution of equation (\ref{farode}) is
\begin{align}
R_{l m}(x) & =A_1\exp (-i D x)  x^l U(l+1,2 l+2,2 i D x)+A_2\exp (-i D x)  x^{-l-1} U(-l,-2 l, 2 i D x),\label{farsol}
\end{align}
where $U(l+1,2 l+2,2 i D x)$ and $U(-l,-2 l, 2 i D x)$ are confluent hyper-geometric functions. To match the solution above with (\ref{nearfarsol}), it is essential to know the small $x$ behavior of the above solution \cite{Brito:2015oca}
\begin{align}
    R_{l m}(x) \sim A_1 x^l+A_2 x^{-l-1}.\label{farnearsol}
\end{align}

\begin{figure}[h]

\includegraphics[width=5in]{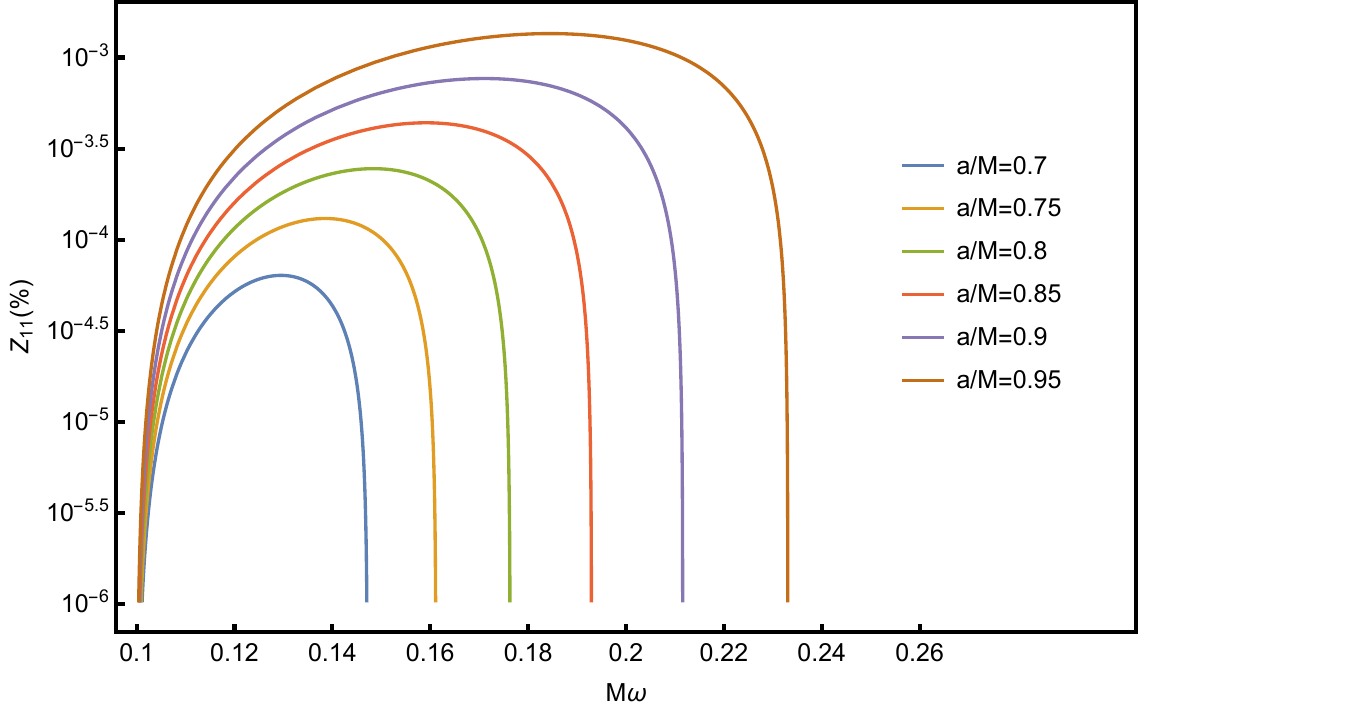}\label{zlma}
\hspace{2mm}
\includegraphics[width=5in]{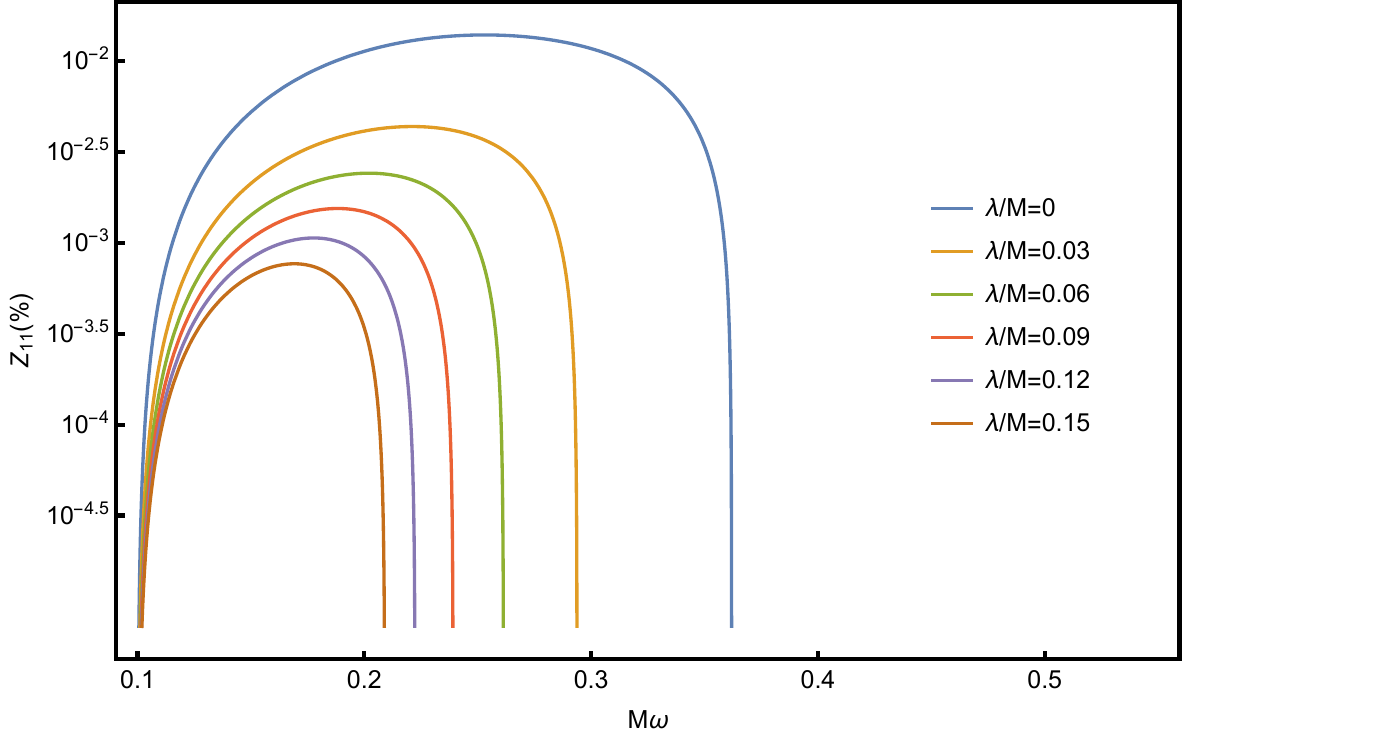}\label{zλ}
\caption{The amplification factor $Z_{11}$ for $l=m=1,\mu M=0.1$. We set parameters $\lambda/M=0.1$ in top panel, parameters $a/M=0.95$ in bottom panel. The top panel shows that the superradiant amplification increases monotonically with the black hole spin $a$. The bottom panel shows that the superradiant amplification effect decreases as the dark matter parameter $\lambda$ increases. Different plot lines in two figures have different frequency ranges, which are limited by the superradiance condition $\mu < \omega < m \Omega_h$.}
\label{z11single}
\end{figure}

\textbf{(iii) Solution in the overlapping region} \quad Now, the two asymptotic solutions Eq.(\ref{nearfarsol}) and Eq.(\ref{farnearsol}) can be matched since they have an overlapping region. At first step, we can get the following coefficient relations,
\begin{align}
A_1 & =C \frac{\Gamma(1-2 i B) \Gamma(2 l+1)}{\Gamma(l+1) \Gamma(l+1-2 i B)} ,\notag\\
A_2 & =C \frac{\Gamma(1-2 i B) \Gamma(-1-2 l)}{\Gamma(-l-2 i B) \Gamma(-l)}.\label{coec}
\end{align}
And we know the solution (\ref{boundary}) takes the following form when $r\rightarrow \infty$,
\begin{align}
    R_{l m} \sim \frac{U_{\infty}\left(r_*\right)}{r} \sim {I} \frac{\exp \left(-i k_{\infty} r_*\right)}{r}+{R} \frac{\exp \left(i k_{\infty} r_*\right)}{r}.\label{solinf}
\end{align}
Then, we need to connect the coefficients $A_1$ and $A_2$ with the coefficients $I$ and $R$ in the asymptotic solution (\ref{boundary}). In order to do that, we expand (\ref{farsol}) at infinity and match with (\ref{solinf}). After some algebra, we obtain the analytical expression for $I$ and $R$ 
\begin{align}
     & {I}=A_1 \frac{(-2 i)^{-l-1} D^{-l} \Gamma(2 l+2)}{k_{\infty} \Gamma(l+1)}+A_2 \frac{(-2 i)^l D^{l+1} \Gamma(-2 l)}{k_{\infty} \Gamma(-l)},\notag\\ & 
    {R}=A_1 \frac{(2 i)^{-l-1} D^{-l} \Gamma(2 l+2)}{k_{\infty} \Gamma(l+1)}+A_2 \frac{(2 i)^l D^{l+1} \Gamma(-2 l)}{k_{\infty} \Gamma(-l)}.\label{IR}
\end{align}
Thus we can get the amplification factor for scalar wave (\ref{Z}) as follows
\begin{align}
    Z_{l m}(\omega)=-4 B D^{2 l+1} \frac{(l !)^4}{((2 l) !)^2((2 l+1) ! !)^2} \times \prod_{k=1}^l\left(1+\frac{4 B^2}{k^2}\right).\label{ZZ}
\end{align}
There will be superradiant phenomena when $Z_{lm}>0$ . In Fig.~\ref{z11single}, we plot $Z_{11}$ for different angular momentum $a$ and PFDM parameter $\lambda$ . \footnote{This equation Eq.(\ref{ZZ}) remains valid for spin values $a \leq M$ if the condition $\mu M<\omega M\ll 1$ is satisfied. }

From the top panel, we observe that the super-radiant amplification factor monotonically increases with the black hole angular momentum $a$. This indicates that the superradiant effect becomes more pronounced as the black hole rotates faster. In the bottom panel, the superradiant amplification weakens gradually as the dark matter parameter increases, which means the dark matter has supressive effect on the black hole superradiance, we will give an explanation about this behavior in the discussion.


\section{Energy extraction from black hole}
\label{energy}
In the last section, we have demonstrated the effect of dark matter density $\lambda$ on the super-radiant amplification factor of scalar waves. While the amplification factor is the most direct quantity to characterize the strength of super-radiance, it is not so straightforward to observe. So in this section, in order to show the effect of dark matter on super-radiance more intuitively, we will investigate other physical quantity which is directly related to super-radiance. It has long been known that super-radiance can be used to extract energy from black hole. Thus in this section, we will further investigate the energy extraction behavior and find the effect of PFDM parameters on the net extracted energy from black hole. 

Firstly, the outgoing energy $\dot{E}_{out}$ measured by an observer at infinity can be obtained by calculating the energy-momentum tensor of the field. For a monochromatic scalar field, the outgoing energy flux is as follows~\cite{Brito:2015oca,Wondrak:2018fza},
\begin{align}
\dot{E}_{out}=\frac{\omega k_{\infty}}{2} |R|^2\label{energy1}.
\end{align}
we leave the derivation of this expression to  Appendix~\ref{A}.  In this section,  we use the massless scalar waves($\mu=0$) to extract energy without loss of generality. The generalization to massive scalar waves will be straightforward. 

It is easy to generalize this to non-monochrome case. In non-monochrome case, we need to integrate the flux contribution of each mode and take into account the Bose-Einstein distribution since we focus on scalar wave. Thus the total extracted energy becomes
\begin{align}
    \dot{E}_{tot}=\int \mathrm{d^{3}k} \frac{\omega k_{\infty}}{2}|R|^2 n(\omega)=\int d\omega \frac{\omega k_{\infty}}{2}|R|^2 \omega^{2}n(\omega) .\label{Eto}
\end{align}
Here, $n(\omega)$ represents the (normalized) distribution function 
\begin{align}
    n(\omega)=\frac{1}{8\pi\zeta(3) T_{tem}^3\left(\exp \left[\omega /T_{tem}\right]-1\right)},
\end{align}
where $8\pi \zeta(3) T_{tem}^3$ is the normalization factor, $\zeta(x)$ is the Riemann zeta function with $\zeta(3)\approx 1.2$. $T_{tem}$ is the temperature of radiation. As $|R|^{2}$ in Eq.(\ref{Eto}) does not depend on $T_{tem}$, the choice of temperature $T_{tem}$ will not affect the result qualitatively but only shift the specific value of extracted energy. However,physically, we should choose $T_{tem} \gg T_{H}$ to ignore the contribution of Hawking radiation.

\begin{figure}[h]
\centering
\includegraphics[width=6in]{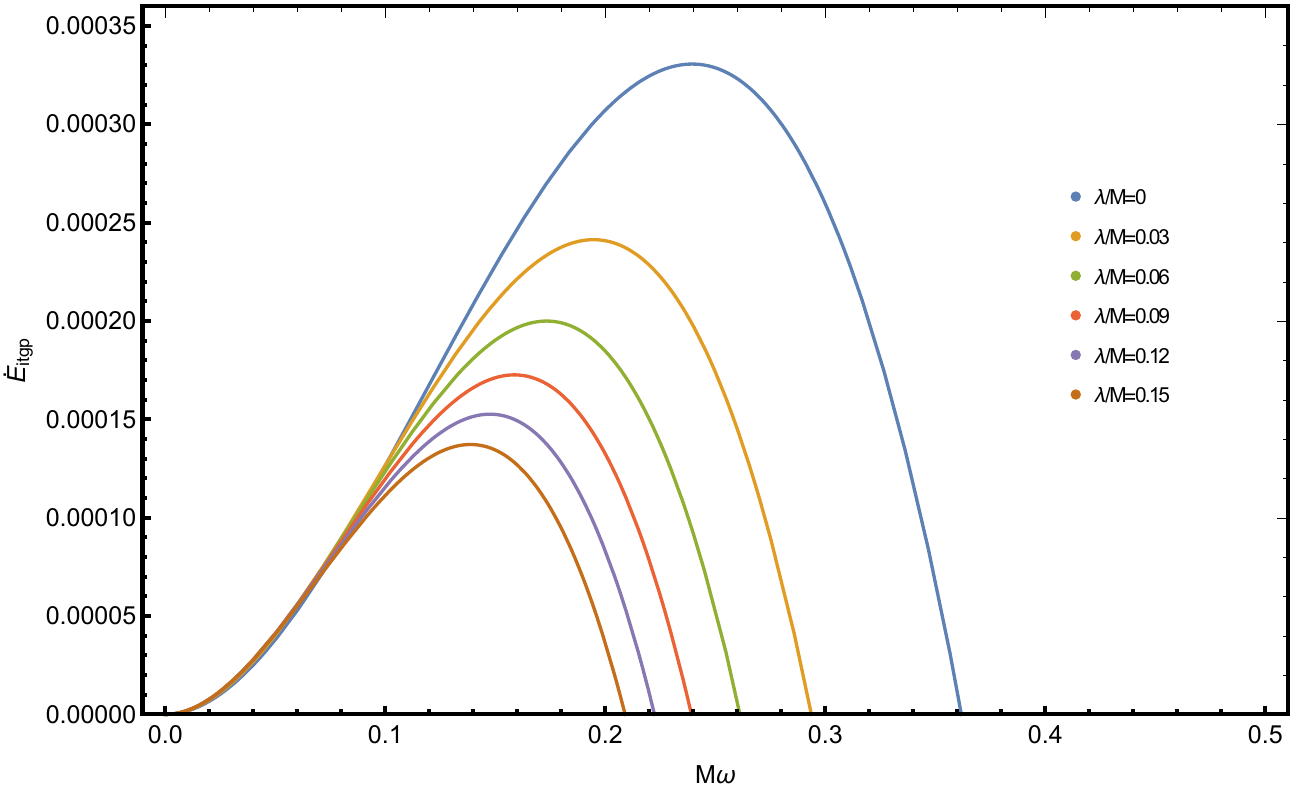}
\caption{The net extracted energy profile $\dot{E}_{itgp}$ at infinity for $l=m=1$ . The spin $a$ are set to $0.95 M$. As seen from this panel, the net extracted energy decreases with increasing dark matter density and is smaller for dark matter black hole compared to Kerr black holes.We choose $T_{tem}=2.7K$ for this plot }
\label{Ep}
\end{figure}

By taking into account the energy of in-going waves, we can obtain the incident energy flux using the same method as obtaining (\ref{energy1}). By subtracting the energy of in-going waves, we can determine the net energy extracted from the black hole via super-radiance as follows
\begin{align}
    \dot{E}_{\text {ptot}}=\int \mathrm{d^{3}k} \frac{\omega k_{\infty}}{2}\left(|R|^2-|I|^2\right) n(\omega)=\int d\omega \frac{\omega k_{\infty}}{2}(|R|^2-|I|^{2}) \omega^{2}n(\omega).
    \label{energyp}
\end{align}
Fig.~\ref{Ep} plots the relation between net extracted energy profile (the integrand in Eq.(\ref{energyp}))and wave frequency. Total net extracted energy is the area of the region between the curve and x-axis. It can be observed that the net energy extraction from PFDM rotating black holes via super-radiance is smaller than that from Kerr black holes , and it decreases further with increasing the dark matter density. This result is consistent with the result of super-radiant amplification factor shown in Sec.\ref{section3}

\section{superradiant instability analysis}
\label{section4}
It has long been known that superradiance can trigger the instability of black hole spacetime \cite{Witek:2012tr}. So there is a natural question about whether this instability can be affected by the presence of dark matter. In this section, we will explore the influence of dark matter parameters on the stability of black holes using the black hole bomb model ~\cite{Press:1972zz}. 

The black hole bomb model puts a mirror outside the black hole to confine and amplify waves through continuous reflections. The presence of mirror is equivalent to impose an effective potential well outside the black hole which is essential to trigger the super-radiant instability.

From equation (\ref{reradialeq}) we have
\begin{align}
    \Delta \frac{d}{d r}\left(\Delta \frac{d R_{l m}(r)}{d r}\right)+ \alpha R_{l m}(r)=0,\label{rr}
\end{align}
where
\begin{align}
   \alpha =(\omega (a^2+r^2)-am)^2-\Delta(r^2\mu^2-2am\omega+a^2\omega^2+l(l+1)).
\end{align}
According to the black hole bomb mechanism, the solution of the radial equation (\ref{rr}) takes the following form
\begin{align}
    R_{l m} \sim \begin{cases}e^{-i(\omega-m \Omega_h) r_*} & \text { as } r \rightarrow r_{h}\left(r_* \rightarrow-\infty\right) \\ \frac{e^{-\sqrt{\mu^2-\omega^2 r_*}}}{r} & \text { as } r \rightarrow \infty\left(r_* \rightarrow \infty\right)\end{cases}
\end{align}
From the equation above, we can deduce the boundary conditions: at the black hole horizon, there exists only a purely ingoing wave, while under the condition $\omega^2<\mu^2$, at spatial infinity, there exists a decaying scalar wave.

Now, with the new radial function
\begin{align}
    \psi_{l m} \equiv \sqrt{\Delta} R_{l m},
\end{align}
the equation (\ref{rr}) becomes 
\begin{align}
    (\frac{d^2}{d^2r}+\omega^2-V_{eff}) \psi_{lm}=0,
\end{align}
where 
\begin{align}
    & \omega^2-V_{eff}=\frac{\alpha+ \beta}{\Delta^2},\\ \notag
    & \beta=\frac{\Delta'(r)^{2}}{4}-\frac{\Delta(r)\Delta''(r)}{2}=\frac{\left(2 M-2 r+\lambda+\lambda \ln \left(\frac{r}{\lambda}\right)\right)^2}{4} -\frac{\left(2-\frac{\lambda}{r}\right)\left(a^2-2 M r+r^2-r \lambda \ln \left(\frac{r}{\lambda}\right)\right)}{2}.
\end{align}
By discarding the terms $\mathcal{O}(1/r^2)$, the effective potential can be approximated as follows
\begin{align}
    V_{eff} \approx \mu^2+\frac{\left(\mu^2-2 \omega^2\right)\left(2 M+\lambda \ln \left(\frac{r}{\lambda}\right)\right)}{r}.\label{va}
\end{align}
If the derivative of the effective potential is asymptotically positive, then there will exist a potential well that can  confine the super-radiant waves and trigger instability. By taking the derivative, we can get the derivative form of equation (\ref{va}) as follows
\begin{align}
    V_{eff}^{\prime}=\frac{\left(2 \omega^2-\mu^2\right)\left(2 M-\lambda(1- \ln \left(\frac{r}{\lambda}\right))\right)}{r^2}.
\end{align}
By demanding $2 M-\lambda(1- \ln \frac{r}{\lambda})>0$, or equivalently $r>\lambda e^{1-\frac{2M}{\lambda}}$. To satisfy condition $V_{eff}^{\prime}>0$, we need to let 
\begin{equation}
    2 \omega^2-\mu^2>0 
\end{equation}
and then we can obtain
\begin{equation}
    \mu<\sqrt{2} \omega 
\end{equation} 
Combining condition $\omega<m \Omega_h$, we can deduce
\begin{align}
    \mu<\sqrt{2}m \Omega_h.
\end{align}
From the above equation, we see that compared to Kerr black hole, the presence of dark matter has no influence on the superradiant instability condition of the black hole in the leadinig order approximation. It is also interesting to compare with behavior in Kiselev black hole case which is reported in Ref.\cite{Khodadi:2021mct}. In this case the metric function $\Delta$ reads
\begin{equation}
    \Delta(r)=r^{2}-2Mr+a^{2}-K r^{1-3\alpha}
\end{equation}
where $K$ is related to dark matter density and $\alpha$ is the state parameter. The author found that for $\alpha=0$ and $\alpha=\frac{1}{3}$ case, the superradiant stability condition is unaffected in the leading order while for $\alpha=-\frac{1}{3}$ the stability condition gets modified. This is because for $\alpha=-1/3$ case, there is a shift in the coefficient of $r^{2}$ term in $\Delta(r)$ expression. As the effective potential depends heavily on the behavior of $\Delta$, thus in the leading order, the effective potential will be largely modified and result in the shift of stability condition by $V'_{eff}>0$ in the asymptotic region. While in our case, from the analysis of effective potential, there is no modification of super-radiant stability condition as the dark matter backreaction of our case is not that large. In this respect, our example resembles the $\alpha=0,\frac{1}{3}$ case in Ref. \cite{Khodadi:2021mct}.

\section{conclusion and disscuion}
\label{section5}
In this work, we investigated the super-radiant amplification effect of rotating black hole surrounded by dark matter.We consider the scattering of the massive scalar field to show this. For the frequency domain where superradiance occurs, the frequency domain shrinks as we increase the dark matter density. This is due to the decrease of angular velocity of black hole when adding the dark matter. For the magnitude of superradiance, we found that as the dark matter parameter increases, the super-radiant amplification factor decreases which means the presence of dark matter will suppress the super-radiance. We also computed the energy extraction behavior from this black hole to further support our results. Finally, we discussed the instability caused by superradiance within the context of the black hole bomb mechanism.We found that the presence of dark matter does not change the mass condition for black hole superradiant instability. 

The suppression effect of dark matter on black hole superradiance can be understood from the black hole spacetime structure. To illustrate this point clearly, we first focus on a much simpler case, Kerr black hole. For Kerr black hole, the existence of ergo-sphere is an important property which is crucial to the energy extraction and superradiance phenomenon. The ergo-sphere describes the difference between the radius of infinite red-shift surface and radius of event horizon. For Kerr black hole, simple computation shows that infinite redshift surface located at
\begin{equation}
    r_{rs}=M+\sqrt{M^{2}-a^{2}\cos^{2}\theta}
\end{equation}
while event horizon located at 
\begin{equation}
    r_{h}=M+\sqrt{M^{2}-a^{2}}
\end{equation}
Thus it can be found that 
\begin{equation}
    \frac{d(r_{rs}-r_{h})}{da}=\frac{a(\sqrt{M^{2}-a^{2}\cos^{2}\theta}-\sqrt{M^{2}-a^{2}}\cos^{2}\theta)}{\sqrt{M^{2}-a^{2}}\sqrt{M^{2}-a^{2}\cos^{2}\theta}}>0
\end{equation}
The inequality follows from $\sqrt{M^{2}-a^{2}\cos^{2}\theta}>\sqrt{M^{2}-a^{2}}>\sqrt{M^{2}-a^{2}}\cos^{2}\theta$.
Thus increase of black hole spin $a$ will enlarge the ergosphere of black hole. This result will also hold when the black hole is slightly shifted from Kerr case. It is also widely accepted that the superradiant amplification factor monotonically increase with the increase of spin, which has been reported in many works \cite{Brito:2015oca,Li:2022hkq}. Thus we can also use this physical intuition to explain the dark matter effect on superradiance. The analytical derivation of the effect of dark matter on ergosphere is difficult. But numerically, as can be seen from Table \ref{ergo}, we can observe that as the PFDM parameter increases, the corresponding  ergo-sphere shrinks. This gives the reason why the superradiant amplification factor and extracted energy decrease as dark matter density $\lambda$ increases. We believe this argument is general and would like to check this in other context in the future.

\begin{table}[h]
    \centering
    \renewcommand\arraystretch{1.2}
    \setlength{\tabcolsep}{1.2pt}
    \begin{tabular}{|c|c|c|c|c|c|}
    
        \hline
        $\lambda/M$& Infinite redshift surface & horizon($r_{h}$) & Ergo-sphere & Ratio & $\Dot{E}_{p11}$  \\
        \hline
         0  & 2   & 1.31225    & 0.68775 & 1.5241  & $6.76136\times 10^{-5} $    \\
       \hline
         0.03  & 2.12785   & 1.52678  & 0.601073 & 1.39369 & $4.0026\times 10^{-5} $     \\
        \hline
         0.06  & 2.21656   & 1.65297    & 0.563589 & 1.34095  & $2.94907\times 10^{-5} $    \\
        \hline
         0.09 & 2.29134 & 1.75209 & 0.539249 & 1.30778 & $2.32864\times 10^{-5} $      \\
        \hline
        0.12 & 2.35734  & 1.83568  & 0.521658 & 1.28418 & $1.91409\times 10^{-5} $      \\
        \hline
        0.15 & 2.41694  & 1.90869   & 0.508249 & 1.26628 & $1.61708\times 10^{-5} $      \\
        \hline
    \end{tabular}
    \caption{Under the condition $a/M=0.95$, the infinite redshift surface, event horizon, ergosphere and ratios corresponding to different parameters $\lambda$, as well as the net energy extraction in the $l=m=1$ mode.We choose $\theta=\frac{\pi}{2}$ to show the difference. The difference between the infinite redshift surface radius and the event horizon radius represents the ergosphere range, while their ratio indicates the relative variation of the ergosphere. The table reflects that the region of ergosphere shrink for increasing dark matter parameter, which is the reason why the pure energy extraction decrease.}
    \label{ergo}
\end{table}


For future research, it is interesting to find more kinds of analytical black hole solutions when dark matter distribution takes different form, especially in the case when dark matter takes Hernquist or NFW form. The progress in this direction will tell us how general the suppressive effect of dark matter on black hole superradiance is. Moreover, as the black hole super-radiance has huge potential to be detected in the future,it is interesting to see if the superradiant properties we found in this paper can be used to search for the galactic dark matter in universe.

\section*{Acknowledgement}
We are grateful to Drs. Shi-Bei Kong, Xiao-kun Yan, Xing-kun Zhang, Xiao-jun Gao and Xiao Liang for interesting and stimulating discussions. Ya-Peng Hu is supported by National Natural Science Foundation of China (NSFC) under grant Nos. 12175105, 11575083, 11565017, 12147175, Top-notch Academic Programs Project of Jiangsu Higher Education Institutions (TAPP).Yu-Sen An is supported by start-up funding No.90YAH23071 and funding No.1018--56XCA2405005 of Nanjing University of Aeronautics and Astronautics (NUAA).

\appendix
\section{Details for the derivation of (\ref{energy1})}\label{A}

In this appendix, we give the detailed derivation of (\ref{energy1}). The Lagrangian density $\mathcal{L}$ of a complex scalar field is 
\begin{align}
    \mathcal{L}=\frac{1}{2} g^{\mu \nu} \partial_{(\mu} \Phi^* \partial_{\nu)} \Phi-\frac{1}{2} \mu^2|\Phi|^2,\label{L}
\end{align}
where $\partial_{(\mu} \Phi^* \partial_{\nu)} \Phi=\frac{1}{2}\left(\partial_\mu \Phi^* \partial_\nu \Phi+\partial_\nu \Phi^* \partial_\mu \Phi\right)$. 
The total energy fluxes at infinity per unit solid angle is given by~\cite{Brito:2015oca}
\begin{align}
    \frac{d^2 E}{d t d \Omega}=\lim _{r \rightarrow+\infty} r^2 T_t^r,\label{dE}
\end{align}
where $T^\mu_\nu$ is the symmetric stress-energy tensor
\begin{align}
    T_\nu^\mu=g^{\mu \lambda} \partial_{(\lambda} \Phi^* \partial_{\nu)} \Phi-\delta_\nu^\mu \mathcal{L}.\label{T}
\end{align}
From the previous context, we know that the scalar field is 
\begin{align}
     \Phi\left(x^\mu\right)=e^{-i \omega t} e^{i m \phi} S_{l m}(\theta) R_{l m}(r),\label{scalar}
\end{align}
and the asymptotic solution at the infinity is 
\begin{align}
    R_{lm}(r)=\frac{I e^{-i k_\infty r}+Re^{i k_\infty r}}{r}.\label{R}
\end{align}
Now, when we substitute (\ref{L}), (\ref{scalar}), and $g^{\mu \nu}$ into (\ref{T}), we can get $T_t^r$ as follows
\begin{align}
    T_t^r=\frac{1}{2}g^{rt}(\partial_t \Phi^* \partial_t \Phi+\partial_t \Phi^* \partial_t \Phi)+\frac{1}{2}g^{rr}(\partial_r \Phi^* \partial_t \Phi+\partial_t \Phi^* \partial_r \Phi) 
   \notag \\+\frac{1}{2}g^{r\theta}(\partial_{\theta} \Phi^* \partial_t \Phi+\partial_t \Phi^* \partial_{\theta}  \Phi)+\frac{1}{2}g^{r\phi}(\partial_{\phi} \Phi^* \partial_t \Phi+\partial_t \Phi^* \partial_{\phi}  \Phi).
\end{align}
Here, as we have deduced from the metric $g^{\mu\nu}$ that $g^{rt}$, $g^{r\theta}$, and $g^{r\phi}$ are all equal to zero, we can remove the corresponding three terms, thus ultimately obtaining the following form,
\begin{align}
    T_t^r=\frac{1}{2}g^{rt}(\partial_t \Phi^* \partial_t \Phi+\partial_t \Phi^* \partial_t \Phi).
\end{align}
We substitute (\ref{R}) into the above equation and can obtain:
\begin{align}
    T_t^r=\frac{1}{2}g^{rr}(-i \omega e^{i\omega t} e^{-im\phi} S^*_{lm}(\theta) R^{*'}_{l m}(r) e^{-i\omega t} e^{i m\phi} S_{l m}(\theta) R_{l m}(r)
   \notag \\+i \omega e^{i\omega t} e^{-i m\phi} S^*_{l m}(\theta) R^*_{l m}(r) e^{-i\omega t} e^{i m\phi} S_{l m}(\theta) R^{'}_{l m}(r) ).\label{Trt}
\end{align}
The terms $R^{*'}_{l m}(r)$ and $R^{'}_{l m}(r)$ denote the derivative with respect to $r$. Then we substitute (\ref{R}) into (\ref{Trt}) and after some calculations and simplification, we can obtain
\begin{align}
    \frac{d E_{out}}{d t }=\frac{\omega k_{\infty}}{2} |R|^2.
\end{align}

\end{document}